\documentclass[prd,a4paper,superscriptaddress,nofootinbib,10pt]{revtex4}
\usepackage{amsfonts}
\usepackage{amssymb}
\usepackage{amsmath}
\usepackage{marvosym}
\usepackage{txfonts}
\usepackage[T1]{fontenc}
\usepackage{mathtools}
\usepackage{lscape}
\usepackage{mathrsfs}
\usepackage{graphicx}
\usepackage{bm}
\usepackage{color}
\usepackage{slashed}
\usepackage{hyperref}
\usepackage{tikz}
\usepackage{pgfplots}
\usepackage{array}
\usepackage{textcomp}
\usepackage{epsfig}
\usepackage{slashed}
\usepackage{comment}
\usepackage{xcolor}
\usepackage{enumerate}
\pgfplotsset{width=10cm}

\newcommand{\dd}{\partial}

\newcommand{\ba}{\begin{eqnarray}}
\newcommand{\ea}{\end{eqnarray}}

\begin{document}
\begin{flushleft}
\texttt{RBI-ThPhys-2026-05}\\
\end{flushleft}

 \title{Geodesic equation in  noncommutative space: a field theory perspective}

\author{Carolina Matt\'e Gregory}
\email{carolina.gregory@unb.br}
\affiliation{Instituto de Fisica, Universidade de Bras\'{\i}lia
70910-900, Brasilia, DF, Brasil}
\affiliation{International Center of Physics C.P. 04667, Bras\'{\i}lia, DF, Brazil}
\author{Tajron Juri\'c }
\email{tjuric@irb.hr}
\affiliation{Rudjer Bo\v{s}kovi\'c Institute, Bijeni\v cka  c.54, HR-10002 Zagreb, Croatia}
\author{Aleksandr Pinzul}
\email{aleksandr.pinzul@gmail.com }
\affiliation{Instituto de Fisica, Universidade de Bras\'{\i}lia
70910-900, Brasilia, DF, Brasil}
\affiliation{International Center of Physics C.P. 04667, Bras\'{\i}lia, DF, Brazil}

\date{\today}

\begin{abstract}
We derive the geodesic equation for point particles propagating in Moyal-type noncommutative spacetimes using a field-theoretic approach based on the quasi-classical limit of the noncommutative Klein-Gordon equation. Starting from a twisted-geometric construction of the covariant Laplace-Beltrami operator, we obtain the noncommutative Hamilton-Jacobi equation and show that all noncommutative effects are absorbed into an effective, position-dependent mass function $M(x)$ appearing in an otherwise standard relativistic dispersion relation. The corresponding particle dynamics then acquires an additional term in the geodesic equation that takes the form of a fixed external force $F_{\text{NC}}^\mu = -\frac{1}{2} g^{\mu\nu}\partial_\nu M^2(x)$, sourced entirely by the quantum nature of spacetime. We compute this effective mass perturbatively up to fourth order in the noncommutativity parameter for a general metric, proving that all odd-order corrections vanish identically. For the specific case of an $(r-\theta)$ twist applied to spherically symmetric backgrounds, we obtain explicit expressions demonstrating that the leading correction to geodesic motion appears at order $\Theta^2$ and is proportional to the probe particle's mass, while massless particles remain unaffected. 
\end{abstract}

\maketitle

\section{Introduction}

\label{sec:intro}

The study of geodesics, defined as extremal paths of point particles in curved spacetime, has been central to general relativity since Einstein's 1915 formulation \cite{Einstein1915}. The geodesic equation:
\begin{equation}
\frac{d^2 x^\mu}{d \tau^2} + \Gamma^\mu_{\alpha \beta} \frac{dx^\alpha}{d \tau} \frac{dx^\beta}{d \tau} = 0,
\label{eq:geodesic}
\end{equation}
not only explained Mercury's perihelion precession \cite{Weinberg1972}, but continues to underpin modern gravitational physics. Nevertheless, various quantum gravity and modified gravity theories predict deviations from classical geodesic motion in extreme regimes.\\

In $f(R)$ gravity \cite{Hu2007,Starobinsky2007}, modifications to Einstein's theory through functions of the Ricci scalar introduce an extra scalar degree of freedom (the "scalaron"), which couples to matter and modifies particle trajectories. Loop quantum gravity \cite{Ashtekar2000} proposes spacetime quantization, leading to quantum-corrected geodesics that may resolve singularities. Semi-classical approaches \cite{Donoghue1994, Dalvit:1998wr} incorporate one-loop quantum corrections to the Schwarzschild metric, producing ``quantum hair'' that alters geodesics near black holes. String-motivated frameworks, like the Randall-Sundrum model \cite{Randall1999}, allow geodesics to leak into extra dimensions, modifying gravitational dynamics at short distances.\\

Particularly intriguing are Lorentz-violating theories \cite{Kostelecky2004}, where preferred frames induce anisotropic dispersion relations, and Horava-Lifshitz gravity \cite{Horava2009}, which achieves renormalizability through anisotropic scaling between space and time at high energies. Quantum gravity phenomenology \cite{Amelino1998,Gacsmere2005} explores energy-dependent dispersion relations leading to ``rainbow geodesics'', where photon paths become energy-dependent, a prediction testable through gamma-ray burst timing delays \cite{Amelino2013, AlvesBatista:2023wqm, CosmoVerseNetwork:2025alb, Addazi:2021xuf}. The propagation of scalar and fermion particles in curved matrix models was studied in \cite{Battista:2022vvl, Battista:2022hqn, Battista:2023glw}.\\

An important geometric perspective emerges in \cite{Pinzul2014}, where geodesic motion in spacetimes with an explicit foliation structure is studied using spectral geometry methods, motivated by Horava-Lifshitz gravity \cite{Horava2009}, the study of  geodesic motion in spacetimes with explicit foliation structure using spectral geometry methods is performed. This approach reveals how the foliation's extrinsic curvature influences particle trajectories, providing a unified description that bridges algebraic and differential geometric viewpoints and, what is important, it can be useful in different generalizations of the geometric setting. \\

A very general, and in some sense model-independent, feature of modified gravity models that try to incorporate some quantum gravitational effects is that at the Quantum Gravity scale (which is typically taken to be the Planck scale, but not necessarily), the classical notion of spacetime as a smooth manifold should be modified to some sort of noncommutative (NC) (or, at least, generalized) geometry. The very general argument in support of this conclusion was given in \cite{Doplicher:1994zv,Doplicher:1994tu}. Motivated by this, noncommutative geometry \cite{Connes1994} is considered as a mathematically rigorous framework for spacetime quantization. An important work by Seiberg and Witten \cite{Seiberg1999} established the relevance of noncommutative geometry in the framework of super-strings and renewed the interest of the community in the subject of noncommutativity. In \cite{Seiberg1999}, it was shown that in a certain so-called decoupling limit the spacetime coordinates on a brane become noncommutative, $[x^\mu,x^\nu] = i\Theta^{\mu\nu}$. This discovery led to very prolific studies in the area of noncommutative (quantum) field theory, see \cite{Douglas:2001ba,Szabo:2001kg} for some earlier reviews. Theories based on the Moyal algebra $[x^\mu,x^\nu] = i\Theta^{\mu\nu}$ also admit extending the notion of spacetime symmetries, including diffeomorphisms, to their ``twisted'' counterparts, see \cite{Wess2007, Aschieri2006} for a review and \cite{Balachandran:2007kv,Balachandran:2007vx} for an alternative approach to twisted QFT and gravity on Moyal space. Recent advances have demonstrated concrete observational implications: Nicolini et al. \cite{Nicolini2006} showed how noncommutative smearing regularizes black hole singularities, in \cite{Ulhoa:2013gfa} the NC corrections to the deviation angle associated with Mercury’s perihelion were found, the corrections to Regge-Wheeler-Zerilli potentials and the gravitational NC quasinormal modes were addressed in \cite{Herceg:2023zlk, Herceg:2023pmc, Herceg:2024vwc}.\\

Several works based on Feynman's early insights \cite{Feynman1993a, tanimura} about deriving equations of motion from canonical commutation relations have been applied to various noncommutative spaces \cite{ncspacegeo}-\cite{ncspacegeo6}. The violation of equivalence principle is reported for noncommutative spaces compatible with the generalized uncertainty principle \cite{Ghosh:2013qra}. A more formal approach to noncommutative geodesics can be found in \cite{Beggs:2018xxd, ref1}.\\

In this work, we present a systematic derivation of the geodesic equation in Moyal-type noncommutative spacetime, combining perturbative techniques with covariant $\star$-product methods. Our approach is a generalization of the one developed in \cite{Pinzul2014} and the results generalize both Feynman's operator-based approach \cite{Feynman1993a} and modern noncommutative gravity theories \cite{Aschieri2006}, while offering new predictions for quantum gravity phenomenology since our main result is that point particles in a noncommutative spacetime obey the usual relativistic dispersion relation, but with an effective position-dependent mass M(x) induced by noncommutativity and strong gravitational fields. We find that all the noncommutative corrections can be entirely encoded in this effective mass, which leads to a geodesic equation with an additional force term.\\

This work is organized as follows: Section II revisits the derivation of the geodesic equation from a field-theoretical standpoint in the commutative setting, emphasizing how the Hamilton–Jacobi formulation naturally emerges from the quasi-classical limit of the Klein–Gordon equation; Section III extends this construction to Moyal-type noncommutative spacetimes, leading to a NC Hamilton–Jacobi equation and a corresponding noncommutative geodesic equation; Section IV develops a perturbative analysis of the effective mass function
$M(x)$, illustrating how noncommutative corrections first appear at second order in the deformation parameter. Finally, Section V discusses the physical consequences of our results. In several appendices, we provide some background material, complementing this with a nonperturbative treatment, proving the reality of the effective mass $M(x)$ and showing that all odd-order corrections vanish, as well as collect the details of some calculations, important for the main part.

\section{From field theory to geodesic equation: commutative case}\label{sec:commutative}

In this section, we briefly review the approach of \cite{Pinzul2014}, which allows us to derive the geodesic equation starting from the formulation of QFT on a curved background. This approach is useful for generalizations to some non-standard geometries, see \cite{Pinzul2014} for details. In Appendix C, we discuss the non-relativistic version of the approach developed below. This should also serve as a motivation for our approach.

The dynamics of a massive field $\Phi$ from the matter sector in quantum field theory on a curved spacetime is described by the Dirac equation. The field that satisfies the Dirac equation also satisfies the Klein-Gordon (KG) equation (which is easily seen by essentially ``squaring'' the Dirac equation). Therefore, our starting point is the KG equation in some curved space (we use $(-+++)$ signature):
\begin{equation}\label{kg}
\left(\square_g +\frac{m^2c^2}{\hbar^2}\right)\Phi=0 ,
\end{equation}
where $\square_g$ is the Laplace-Beltrami operator for a fixed curved background metric $g_{\mu\nu}$ given by
\begin{equation}\label{lb}
\square_g\Phi=\frac{-1}{\sqrt{g}}\partial_{\mu}(\sqrt{g}g^{\mu\nu}\partial_\nu \Phi),
\end{equation}
and $m$ is the mass of the field. Since we are interested in the geodesic motion, which is defined for a point-like classical particle, we restrict ourselves to the one-particle sector, i.e., we interpret $\Phi$ as a wave function and study it in the quasi-classical approximation. This amounts to writing $\Phi$ in the following form
\begin{equation}\label{ansatz}
\Phi=A \exp\left(\frac{i}{\hbar}S\right),
\end{equation}
where $A$ and $S$ admit asymptotic $\hbar$-expansions
\begin{equation}\label{asympt}
A=A_0+\hbar A_1+..., \quad S=S_0+\hbar S_1+...\ .
\end{equation}
Plugging  \eqref{ansatz} into \eqref{kg} and using \eqref{asympt}, we get to the leading order in  $\hbar$, i.e., in the quasi-classical limit, the equation for $S_0$,
\begin{equation}\label{hj}
g^{\mu\nu}(\partial_{\mu}S_0)( \partial_{\nu}S_0)+m^2c^2=0 .
\end{equation}
The equation \eqref{hj} is nothing but the relativistic Hamilton-Jacobi equation for a point particle in a curved spacetime. Therefore, we can interpret $S_0$ as the classical action, i.e., Hamilton's principal function, and by using the definition of the canonical momentum
\begin{equation}
p_\mu=\partial_{\mu}S_0 ,
\end{equation}
and obtain a relativistic dispersion relation\footnote{Notice that we could have started with the dispersion relation \eqref{dispersion_com}, which is very natural in the commutative case. Instead, we intend to use this construction in the noncommutative case, where the dispersion relation is not that obvious. Therefore we start with the relativistic equation \eqref{kg}, which has a natural analog in the noncommutative framework.}
\begin{equation}\label{dispersion_com}
g^{\mu\nu}p_{\mu}p_{\nu}+m^2c^2=0.
\end{equation}
To analyze the dynamics stemming from this relativistic dispersion relation, one interprets this as Hamiltonian constraint
\begin{equation}
H=g^{\mu\nu}p_{\mu}p_{\nu}+m^2c^2=0 ,
\end{equation}
which leads to the Hamiltonian equations of motion
\begin{equation}
\dot{x}^\mu=N(\tau)\frac{\partial H}{\partial p_\mu}, \quad   \dot{p}_\mu=-N(\tau)\frac{\partial H}{\partial x^\mu} ,
\end{equation}
where the derivative is with respect to some affine parameter $\tau$ and $N(\tau)$ is an arbitrary function of $\tau$, the so-called lapse function, that reflects the time reparametrization freedom. Notice that
\begin{equation}
\dot{x}^{\mu}p_\mu=-2N(\tau)m^2c^2, \quad \dot{x}^{\mu}\dot{x}_\mu=-4N^2(\tau)m^2c^2 ,
\end{equation}
which leads to
\begin{equation}
mcN(\tau)=\frac{1}{2}\sqrt{-g_{\mu\nu}\dot{x}^\mu\dot{x}^\nu}.
\end{equation}
Fixing the reparametrization symmetry by choosing the gauge $N(\tau)=\frac{1}{2}$ corresponds to taking $\tau$ to be the proper time. Then our dynamics reduces to
\begin{equation}\begin{split}\label{set}
&g^{\mu\nu}p_{\mu}p_{\nu}+m^2c^2=0 ,\\
&\ \  \  \ \dot{x}^\mu=g^{\mu\nu}p_\nu ,\\
&\dot{p}_\mu=-\frac{1}{2}\frac{\partial g^{\alpha \beta}}{\partial x^\mu}p_\alpha p_\beta .
\end{split}\end{equation}
Solving for $p_\mu$, \eqref{set} immediately leads to the geodesic equation, where as usual, $g_{\alpha \mu}g^{\mu\beta}=\delta^{\beta}_{\alpha}$,
\begin{equation}
\ddot{x}^\mu+\Gamma^{\mu}_{\ \alpha\beta}\dot{x}^\alpha\dot{x}^\beta =0 ,
\end{equation}
and
\begin{equation}
\Gamma^{\mu}_{\ \alpha\beta}=\frac{1}{2}g^{\mu\nu}(\partial_\alpha g_{\nu \beta}+\partial_\beta g_{\nu \alpha}-\partial_\nu g_{\alpha \beta})
\end{equation}
are the standard Christoffel symbols. Thus, we have illustrated how to derive a geodesic equation of a point particle starting from the field theory perspective \eqref{kg}.

\section{From noncommutative field theory to noncommutative geodesics}

As illustrated in the previous section, one can obtain the geodesic equation starting from the KG equation. In this section, we apply the same method to the noncommutative generalization of the KG equation with the aim of obtaining the NC generalization of the geodesic equation. One of the approaches to the NC KG equation is based on the well-known formalism of NC differential geometry \cite{Herceg:2023zlk,
Herceg:2023pmc, Aschieri:2005zs, Aschieri:2005yw, Aschieri:2009qh, schenkel, Aschieri:2017ost, Aschieri:2020yft}. Since this is a crucial (starting) point of our approach, for the interested reader, in Appendix A, we have outlined the derivation of NC KG equation with all the details. Here we simply motivate the origin of the NC generalization of Laplace-Beltrami operator $\hat{\square}_g$ and proceed with the derivation of the NC Hamilton-Jacobi equation, NC dispersion relation and finally NC geodesic equation.

As it is explained in detail in Appendix A, our approach to KG equation is based on the geometric definition of the covariant Laplace-Beltrami operator \eqref{dAl}
\ba
\Box_g\Phi = * d\! *\! d \Phi ,
\ea
which admits a very natural noncommutative generalization \eqref{NCdAl}
\begin{equation}
\hat{\square}_g\Phi \equiv *^\mathcal{F} d*^{\mathcal{F}}d\Phi .
\end{equation}
This, in turn, leads to the NC KG equation written in terms of the \textit{classical} metric and $\star$-product
\begin{equation}\label{NCKG}
\left(\hat{\square}_g + \frac{m^2 c^2}{\hbar^2}\right)\Phi = \frac{1}{\sqrt{g}}\star \left[\partial_\nu \left[\left(\sqrt{g}~g^{\mu\nu}\right)\star \partial_\mu\Phi\right]\right] + \frac{m^2 c^2}{\hbar^2}\Phi,
\end{equation}
where the $\star$-product between any two functions $f$ and $g$ can be written as (for more details, see Appendix A)
\begin{equation}
f\star g=fg+\frac{i}{2}\Theta^{\mu\nu}\partial_{\mu}f\partial_{\nu}g+\mathcal{O}(\Theta^2).
\end{equation}
In the commutative case, there are several equivalent ways of obtaining the KG equation. It is an interesting and important question whether their NC generalizations (if exist!) will lead to the same noncommutative KG equation \eqref{NCKG}. We do \textit{not} address this question in the present work, but plan on addressing it elsewhere in the future.\\

There seems to be a lot of ambiguity in the choice of the quasi-classical representation for $\Phi$. Our goal is not to construct a fully noncommutative notion of particle trajectory, but rather to determine how noncommutativity modifies the trajectories reconstructed within the ordinary commutative spacetime description, i.e. from the point of view of what we should interpret as a commutative observer. For this reason we employ the standard, commutative, quasi-classical ansatz \eqref{ansatz}, $\Phi=A e^{iS/\hbar}$. A fully noncommutative WKB analysis based on a $\star$-exponential would correspond to a different notion of quasi-classical limit and may contain additional noncommutative effects. Investigating such a construction and its interpretation is beyond the scope of the present work.  Thus, for our purposes, we continue using the ansatz \eqref{ansatz} and taking the quasi-classical limit (i.e., following exactly the same procedure as in the commutative case from Section II), we obtain the NC generalization of the Hamilton-Jacobi equation


\begin{equation}\label{nchj}
\frac{1}{\sqrt{g}}\star\left[\left(\sqrt{g}g^{\alpha\beta}\right)\star\left(\partial_\alpha S_0\ \partial_\beta S_0\right)\right]+m^2c^2=0 .
\end{equation}

In order to extract the NC generalization of the dispersion relation, we first postulate, as in the commutative case, $p_\mu=\partial_\mu S_0$, then $\star$-multiply from the left \eqref{nchj} with $\sqrt{g}_\star $, which is an algebraic $\star$-inverse of $\frac{1}{\sqrt{g}}$, i.e.,
\begin{equation}\label{sgstar}
\sqrt{g}_\star \star \frac{1}{\sqrt{g}}=1,
\end{equation}
and finally obtain
\begin{equation}\label{medju}
(\sqrt{g}g^{\alpha\beta})\star(p_\alpha p_\beta)+m^2c^2\sqrt{g}_\star=0 .
\end{equation}
It is important to notice now that the canonical momentum $p_\mu$ lives in the cotangent sector of the full phase space in the (classical) Hamiltonian mechanics, therefore the $\star$-products between an arbitrary function $F(x)$ and momentum reduce to point-wise multiplication, since the $\star$-product is made of action of the tangent vector fields, i.e., $F(x)\star p_\mu=F(x)p_\mu$. This enables us to drop the $\star$-product in \eqref{medju} and, by dividing the whole equation by $\sqrt{g}$, we finally obtain the NC dispersion relation
\begin{equation}
g^{\mu\nu}p_{\mu}p_{\nu}+M^2(x)=0 ,
\end{equation}
where we introduced an effective mass function, $M(x)$
\begin{equation}\label{M}
M^2(x)=m^2c^2\frac{\sqrt{g}_\star}{\sqrt{g}}.
\end{equation}

We emphasize that the explicit form of the effective mass function $M(x)$ is model-dependent and follows from the particular noncommutative Klein-Gordon operator adopted in this work. Different noncommutative formulations may lead to different corrections to the corresponding Hamilton-Jacobi equation and, consequently, to different expressions for $M(x)$. Nevertheless, the quasi-classical procedure relating the Hamilton-Jacobi equation to particle dynamics is independent of the specific noncommutative model. In this sense, while the detailed form of the correction is model-dependent, the appearance of modifications to the effective particle dynamics induced by noncommutativity is expected to be a more general feature. Another point worth emphasizing is that the effective mass function $M(x)$ should not be interpreted as an independent dynamical field or as a new propagating degree of freedom. Rather, it is a derived geometric quantity, completely determined by the background geometry and the noncommutative deformation. In this sense, $M(x)$ plays the role of an effective (local) parameter encoding noncommutative corrections to particle motion rather than introducing additional dynamics.\\ 

Some comments are in order. Firstly, we see that the difference between the commutative and NC dispersion relation is only in the mass term. Namely, all the NC effects in the NC dispersion relation are absorbed in the effective mass function $M(x)$. Thus, from a physical point of view, the noncommutative dispersion relation of a point particle looks like the commutative one with the position-dependent mass. Secondly, the effective mass function $M(x)$ is completely determined by the metric (or, more precisely, by the volume form) and the type of noncommutativity. Therefore, technically, the key object to calculate is the algebraic $\star$-inverse $\sqrt{g}_\star$ defined by \eqref{sgstar}. This will be the central goal of later sections.\\

To finish the derivation of the NC geodesic equation, we use the same constraint analysis as in Section \ref{sec:commutative}, with the only difference that now $m^2c^2\mapsto M^2(x)$ and the position dependence of the effective mass function results in an additional term in the equation of motion, which looks like an external force\footnote{The effective mass term in the dispersion relation and external force term in the geodesic equation can also be found in the  Bohmian interpretation of relativistic quantum mechanics \cite{Nikolic:2005ds, Nikolic:2012wj}.}
\begin{equation}\label{ncgeo}
\ddot{x}^\mu+\Gamma^{\mu}_{\ \alpha\beta}\dot{x}^\alpha\dot{x}^\beta =-\frac{1}{2}g^{\mu\nu}\frac{\partial M^2}{\partial x^\nu} .
\end{equation}
We see that noncommutativity can be effectively described by a fixed external force $F^{\mu}_{NC}=-\frac{1}{2}g^{\mu\nu}\partial_\nu M^2(x)$. Therefore, the trajectory of a point particle moving in a curved noncommutative space would deviate from the standard geodesic due to this extra force that is sourced by the quantum nature of spacetime. To illustrate the NC effect of the external force $F_{NC}$ we will work out the Newtonian limit of the NC geodesic equation \eqref{ncgeo}.\\

The Newtonian limit is defined by 3 conditions \cite{Carroll:2004st}:

\begin{enumerate}
\item Particles move slowly, i.e. $\frac{d x^i}{d\tau}\ll\frac{dx^0}{d\tau}$.
\item The gravitational field is weak and can be considered as a perturbation of the flat space, i.e. $g_{\mu\nu}=\eta_{\mu\nu}+h_{\mu\nu}$, where $||h_{\mu\nu}||\ll 1$.
\item The gravitational field is essentially static, i.e., $||\partial_t h||\ll ||\nabla h||$.
\end{enumerate}

Taking the above into account, the NC geodesic equation \eqref{ncgeo} reduces to
\begin{equation}
\ddot{x}^0=-\frac{1}{2}g^{0\nu}\frac{\partial M^2}{\partial x^\nu}, \quad \ddot{x}^{i}+\frac{1}{2}\frac{\partial h^{00}}{\partial x_i}=-\frac{1}{2}g^{ii}\frac{\partial M^2}{\partial x^i} ,
\end{equation}
which, after using the Schwarzschild metric and assuming $\frac{\partial M}{\partial x^0}=0$, reduces to
\begin{equation}\begin{split}
\ddot{r}+G\frac{M_{BH}}{r^2}&=-\frac{1}{2}\left(1-\frac{2M_{BH}}{r}\right)\frac{\partial M^2}{\partial r} ,\\
\ddot{\theta}+G\frac{M_{BH}}{r^3}\theta&=-\frac{1}{2r^2}\frac{\partial M^2}{\partial \theta} ,\\
\ddot{\varphi}+G\frac{M_{BH}}{r^3}\varphi&=-\frac{1}{2r^2\sin\theta}\frac{\partial M^2}{\partial \varphi} .\\
\end{split}\end{equation}
As we will see in the next section, the effective mass function $M(x)$ can be written as 
\begin{equation}\label{M2}
M^2 = m^2 c^2 \frac{\sqrt{g_\star}}{\sqrt{g}} = m^2 c^2 \left( 1 + \frac{1}{2} \Theta^{\mu_1 \nu_1} \Theta^{\mu_2\nu_2} \left(\partial_{\mu_1} \partial_{\mu_2} \sqrt{g}\right) \left(\partial_{\nu_1} \partial_{\nu_2} \frac{1}{\sqrt{g}}\right) + \mathcal{O} (\Theta^4) \right).
\end{equation}
Therefore, we see that the Newtonian approximation ``feels'' the noncommutativity only if some components of the curvature, $R^{\mu}_{\ \nu\rho\sigma}$, are ``significant'' on the Quantum Gravity scale or, more concretely, $R\Theta$ (or rather its gradient) should not be too small.\footnote{In fact, one has
\ba
\partial_\mu \partial_\nu \sqrt{g} = \sqrt{g} \left(\partial_\mu \Gamma^{\rho}_{\rho\nu} + \Gamma^{\sigma}_{\sigma\mu} \Gamma^{\rho}_{\rho\nu} \right),\ \ \partial_\mu \partial_\nu \frac{1}{\sqrt{g}} = -\frac{1}{\sqrt{g}} \left(\partial_\mu \Gamma^{\rho}_{\rho\nu} - \Gamma^{\sigma}_{\sigma\mu} \Gamma^{\rho}_{\rho\nu} \right). \nonumber
\ea
Comparing this to $R^{\mu}_{\ \nu\rho\sigma} = \partial_\rho \Gamma^{\mu}_{\ \sigma\nu} - \partial_\sigma \Gamma^{\mu}_{\ \rho\nu} + \Gamma^{\mu}_{\ \rho\lambda}\Gamma^{\lambda}_{\ \sigma\nu} - \Gamma^{\mu}_{\ \sigma\lambda}\Gamma^{\lambda}_{\ \rho\nu}$, leads to the estimate $M^2 = m^2 c^2 \left( 1 + \mathcal{O}(R^2\Theta^2) \right).$ Here $R$ is not necessarily the Ricci scalar, but rather a ``scale of curvature'', or more precisely the scale where $\frac{\partial^2 g_{\mu\nu}}{\partial x^\alpha \partial x^\beta}$ or the gradient of the Christoffel symbol becomes relatively large. \label{footnote}} Note that here $R$ cannot be just the scalar curvature, because, for example, $R=0$ for Schwarzschild solution, while the analysis in the next section shows that the nontrivial noncommutative corrections at this order do exist. The condition that the gradient $\nabla h$ changes significantly on the $\Theta$ scale typically indicates a very strong gravitational field for which the Newtonian limit is not applicable. Therefore, the only hope to probe the noncommutativity is to work with a relativistic system.

\section{The effective mass function $M(x)$: perturbative analysis}

Looking at \eqref{M}, it is clear that the main technical problem is to determine the function $\sqrt{g}_\star$ that has the property of being the algebraic $\star$-inverse of $\frac{1}{\sqrt{g}}$, i.e., to find $\sqrt{g}_\star$ from the condition
\begin{equation}\label{sgstar1}
\sqrt{g}_\star \star \frac{1}{\sqrt{g}}=1
\end{equation}
for a given classical metric with determinant $g$ and fixed noncommutativity determined by $\Theta^{\mu\nu}$. In order to calculate $\sqrt{g}_\star$, we assume that it exists as an element of $\mathcal{A}_\star$ (see Appendix A for more details on $\star$-product algbera). Therefore, it can be expanded as a power series in the deformation parameter $\Theta$, and can be symbolically  written as
\begin{equation}\label{series}
\sqrt{g}_\star=\sum^{\infty}_{n=0} F_n ,
\end{equation}
where $F_n = \Theta^{\mu_1 \nu_1} \cdots \Theta^{\mu_n \nu_n} f_{\mu_1 \cdots \mu_n \nu_1 \cdots \nu_n}$, with $f$'s being elements of $\mathcal{A} = \mathcal{C}^{\infty}(\mathcal{M})$. Using \eqref{series} and the definition of the Moyal $\star$-product \eqref{Moyal_star}, we rewrite \eqref{sgstar1} as
\ba\label{series1}
1 &=& \sum\limits_{n=0}^{\infty} \frac{1}{n!}\left(\frac{i}{2}\right)^n \Theta^{\mu_1 \nu_1} \cdots \Theta^{\mu_n \nu_n} \left( \partial_{\mu_1} \cdots \partial_{\mu_n} \sqrt{g}_\star \right) \left( \partial_{\nu_1} \cdots \partial_{\nu_n} \frac{1}{\sqrt{g}} \right) = \nonumber \\
&=& \sum\limits_{m=0}^{\infty} \sum\limits_{s=0}^{m} \frac{1}{(m-s)!}\left(\frac{i}{2}\right)^{m-s} \Theta^{\mu_1 \nu_1} \cdots \Theta^{\mu_{m-s} \nu_{m-s}} \left( \partial_{\mu_1} \cdots \partial_{\mu_{m-s}} F_s \right) \left( \partial_{\nu_1} \cdots \partial_{\nu_{m-s}} \frac{1}{\sqrt{g}} \right).
\ea
Except for the zeroth order ($m=0$), which must be equal to $1$, all the other orders ($m>0$) should be identically equal to zero. Below, we calculate $\sqrt{g}_\star$, and, as the consequence, $M^2$, up to the fourth order included. In Appendix B, we prove as a general result that all the odd orders must vanish, and demonstrate this explicitly for the first order.\footnote{In Appendix B, we present a slight modification of the calculation done in this section. There we adapt to $2$-dimensional non-commutativity (meaning that only 2 of $d$ dimensions will be non-commutative) and to the specific case of a factorized $\sqrt{g}$. That was done not only to double check the result of this section, but also to see if there is a chance (not yet realized) to find a non-perturbative answer.}\\

In the \textbf{zeroth order}, we are basically dealing with no NC correction, so $F_0=\sqrt{g}$, i.e., $\sqrt{g}_\star =\sqrt{g} + \mathcal{O}(\Theta)$. Up to the \textbf{first order} in $\Theta$, \eqref{series} reduces to $\sqrt{g}_\star=\sqrt{g}+F_1+\mathcal{O}(\Theta^2)$ and the condition \eqref{series1} gives us, for $m=1$,
\begin{equation}
0 =\frac{i}{2}\Theta^{\mu\nu}(\partial_{\mu}F_0)\left(\partial_\nu\frac{1}{\sqrt{g}}\right) + F_1 \frac{1}{\sqrt{g}} \equiv \frac{i}{2}\Theta^{\mu\nu}(\partial_{\mu}\sqrt{g})\left(\partial_\nu\frac{1}{\sqrt{g}}\right) + F_1 \frac{1}{\sqrt{g}},
\end{equation}
which trivially gives $F_1 = 0$ due to $\Theta^{\mu\nu}(\partial_{\mu}\sqrt{g})\left(\partial_\nu\frac{1}{\sqrt{g}}\right) = 0$. This is true in general, namely, for any two functions, $f=f(g)$ and $h=h(g)$, one has $\Theta^{\mu\nu}\partial_{\mu}f \partial_\nu h = 0$. This agrees with the general result from Appendix B. It is important to note that the vanishing of the linear order (as well as all the odd orders, see Appendix B) is obtained for an arbitrary metric and for a generic Moyal space, supporting previous results in the literature indicating that NC corrections to gravity on Moyal type space always appear at $\Theta^2$ order \cite{Mukherjee:2006nd, Aschieri:2005yw, Alvarez-Gaume:2006qrw, Calmet:2006iz}.\\

Next, up to the \textbf{second order}, \eqref{series} is $\sqrt{g}_\star=\sqrt{g}+F_2+\mathcal{O}(\Theta^3)$, the condition \eqref{series1} for $m=2$ is now
\begin{equation}
0 =\frac{1}{2!}\left(\frac{i}{2}\right)^2 \Theta^{\alpha\beta}\Theta^{\gamma\delta}(\partial_{\alpha}\partial_{\gamma}F_0)\left(\partial_{\beta}\partial_{\delta}\frac{1}{\sqrt{g}}\right) + F_2 \frac{1}{\sqrt{g}} \equiv \frac{1}{2!}\left(\frac{i}{2}\right)^2 \Theta^{\alpha\beta}\Theta^{\gamma\delta} (\partial_{\alpha}\partial_{\gamma}\sqrt{g})\left(\partial_{\beta}\partial_{\delta}\frac{1}{\sqrt{g}}\right) + F_2 \frac{1}{\sqrt{g}},
\end{equation}
which leads to
\begin{equation}\label{2nd}
F_2=-\frac{\sqrt{g}}{2} \left(\frac{i}{2}\right)^2 \Theta^{\alpha\beta}\Theta^{\gamma\delta} (\partial_{\alpha}\partial_{\gamma}\sqrt{g})\left(\partial_{\beta}\partial_{\delta}\frac{1}{\sqrt{g}}\right).
\end{equation}
Note that this expression is also given for any metric and a general Moyal space.\\

As we already mentioned above, the \textbf{third order}, i.e., $F_3$, must be zero, which is also straightforward to verify by explicitly calculating $F_3$ using \eqref{series}.\\

Up to the \textbf{fourth order}, \eqref{series} is $\sqrt{g}_\star=\sqrt{g}+F_2+F_4+\mathcal{O}(\Theta^6)$ and the condition \eqref{series1}, $m=4$, is now
\ba
0 = \frac{1}{4!} \left(\frac{i}{2}\right)^4 \Theta^{\mu_1 \nu_1}\Theta^{\mu_2 \nu_2}\Theta^{\mu_3 \nu_3}\Theta^{\mu_4\nu_4} (\partial_{\mu_1}\partial_{\mu_2}\partial_{\mu_3}\partial_{\mu_4}F_0) (\partial_{\nu_1}\partial_{\nu_2}\partial_{\nu_3}\partial_{\nu_4}\frac{1}{\sqrt{g}}) + \frac{1}{2!} \left(\frac{i}{2}\right)^2 \Theta^{\alpha\beta}\Theta^{\gamma\delta} (\partial_{\alpha}\partial_{\gamma} F_2) \left(\partial_{\beta}\partial_{\delta}\frac{1}{\sqrt{g}}\right) ,
\ea
which gives us the expression for $F_4$:
\begin{equation}\label{4th}
F_4=-\sqrt{g}\left(\frac{1}{4!}\left(\frac{i}{2}\right)^4\Theta^{\mu_1 \nu_1}\Theta^{\mu_2 \nu_2}\Theta^{\mu_3 \nu_3}\Theta^{\mu_4\nu_4}(\partial_{\mu_1}\partial_{\mu_2}\partial_{\mu_3}\partial_{\mu_4}\sqrt{g}) (\partial_{\nu_1}\partial_{\nu_2}\partial_{\nu_3}\partial_{\nu_4}\frac{1}{\sqrt{g}})+ \frac{1}{2}\left(\frac{i}{2}\right)^2\Theta^{\alpha\beta}\Theta^{\gamma\delta}(\partial_{\alpha}\partial_{\gamma} F_2) \left(\partial_{\beta}\partial_{\delta}\frac{1}{\sqrt{g}}\right)\right) ,
\end{equation}
where $F_2$ is given by \eqref{2nd}. As before, the result is valid for an arbitrary metric and a general Moyal deformation. Now, we specify to the case of the so-called $(r-\theta )$ twist, that can be conveniently written as
\begin{equation}\label{rttwist}
\mathcal{F} = \exp(-i\lambda (\partial_r\otimes \partial_\theta - \partial_\theta\otimes \partial_r) ),
\end{equation}
where the only nontrivial components of $\Theta^{\mu\nu}$ are $\Theta^{r\theta} = -\Theta^{\theta r} = 2\lambda$, and $\lambda$ is of the dimension of length and represents the NC or Quantum Gravity scale. This is, in some sense, a minimal (non-Killing) twist to produce a non-trivial NC deformation of a spherically symmetric spacetime. To be even more specific, we take the 4-dimensional metric to be
\ba\label{metric_spher}
ds^2 = -f(r)dt^2 + \frac{1}{f(r)}dr^2 + r^2 (d\theta^2 + \sin^2 \theta \, d\phi^2),
\ea
i.e., $\sqrt{g} = r^2 \sin\theta$. Then we have from $F_0 = \sqrt{g}$, \eqref{2nd} and \eqref{4th}
\ba
F_2 = -2 \lambda^2 \frac{1}{\sin\theta} ,\ \ F_4 = 4\lambda^4 \frac{1}{r^2 \sin\theta},
\ea
leading to
\ba\label{g_expension}
\sqrt{g}_\star = r^2 \sin\theta \left(1-\frac{2\lambda^2}{r^2\sin^2\theta}+\frac{4\lambda^4}{r^4 \sin^2 \theta} + \mathcal{O}(\lambda^6) \right),
\ea
in complete agreement with the result \eqref{4th_xy} from Appendix B. For $M^2 (r,\theta )$, we have, up to the fourth order, the following result
\begin{equation}
M^2=m^2c^2\left(1-\frac{2\lambda^2}{r^2\sin^2\theta}+\frac{4\lambda^4}{r^4 \sin^2 \theta} + \mathcal{O}(\lambda^6) \right)
\end{equation}
or, taking into account the specific form of our metric,
\begin{equation}
M^2=m^2c^2\left(1-\frac{2\lambda^2 r^2}{g}+\frac{4\lambda^4}{g}  + \mathcal{O}(\lambda^6)\right).
\end{equation}
These formulas look very suggestive of the existence of a ``nice'' function, such that \eqref{g_expension} represents the first terms of its expansion, although we have not yet been able to demonstrate this.

\section{Discussion and concluding remarks}

Upon combining \eqref{ncgeo} with \eqref{2nd}, we get the general expression for the geodesic equation with leading non-trivial noncommutative corrections:
\begin{equation}\label{ncgeo1}
\boxed{\ddot{x}^\mu+\Gamma^{\mu}_{\ \alpha\beta}\dot{x}^\alpha\dot{x}^\beta =-\frac{m^2 c^2}{4}g^{\mu\nu} \partial_\nu \left[ \Theta^{\mu_1 \nu_1} \Theta^{\mu_2\nu_2} \left(\partial_{\mu_1} \partial_{\mu_2} \sqrt{g}\right) \left(\partial_{\nu_1} \partial_{\nu_2} \frac{1}{\sqrt{g}}\right)\right] + \mathcal{O} (\Theta^4).}
\end{equation}
Note that, in our approach, noncommutative corrections are essentially ``isodeterminant'', meaning that all the geometries with the same determinant of the spherically symmetric metric given in \eqref{metric_spher} will have the same effective mass (and this is true to \textit{all} orders), and the only dependence on the metric is due to $g^{\mu\nu}$ on the right hand side of \eqref{ncgeo1}.\footnote{We expect that the right hand side of \eqref{ncgeo1} can be written in a more geometric way through some components of the Riemann tensor, see footnote \ref{footnote}, although we have not yet found such an expression for it.\label{footnote1}} For instance, in the case of the metric \eqref{metric_spher}, the corrected geodesics take the form (to the leading order):
\[
\begin{aligned}
&\ddot t + \frac{f'(r)}{f(r)}\,\dot t\,\dot r = 0, \\[6pt]
&\ddot r + \tfrac{1}{2}f(r)f'(r)\,\dot t^{2} - \frac{f'(r)}{2f(r)}\,\dot r^{2} - r f(r)\!\left(\dot\theta^{2}+\sin^{2}\theta\,\dot\phi^{2}\right) = \frac{2m^2 c^2 f(r)}{r^3 \sin^2 \theta}\lambda^2 + \mathcal{O}(\lambda^4), \\[6pt]
&\ddot\theta + \frac{2}{r}\,\dot r\,\dot\theta - \sin\theta\cos\theta\,\dot\phi^{2} = \frac{2m^2 c^2 \cos\theta}{r^4 \sin^3 \theta}\lambda^2 + \mathcal{O}(\lambda^4), \\[6pt]
&\ddot\phi + \frac{2}{r}\,\dot r\,\dot\phi + 2\cot\theta\,\dot\theta\,\dot\phi = 0.
\end{aligned}
\]
So, the only dependence on $f(r)$ is due to the explicit $g^{\mu\nu}$ dependence. Note that, as it should be clear from the discussion in Appendix B, the equations for $t$ and $\phi$ will remain unmodified to all orders.\\

As we discussed at the end of Section III, the leading non-trivial corrections to $M^2$ are of the symbolic form $R^2 \Theta^2$, which shows that the effective noncommutativity should be typically ``enhanced'' by strong gravity. Although in some highly symmetric cases, like the one of \eqref{metric_spher}, this effect could be masked, still this can be thought as some sort of UV/IR effect: while noncommutativity is clearly a UV phenomenon of Quantum Gravity, the enhancement is due to GR, which is an IR regime of \textit{the same} Quantum Gravity theory.\\

Another important noteworthy consequence is that, as an immediate consequence of equation \eqref{ncgeo}, there is a violation of the weak equivalence principle, since the noncommutative force depends on the particle mass through the effective mass function $M(x)$. However, the leading correction scales as $R^2\Theta^2$. If the noncommutativity scale is of the order of the Planck scale, $\Theta \sim \ell_{\rm Pl}^2$, the effect is strongly suppressed whenever the characteristic curvature radius is much larger than the Planck length. Consequently, any deviation from geodesic motion is expected to be completely negligible in ordinary astrophysical or laboratory settings and may become relevant, as is the case with essentially all effects due to noncommutativity, only in regimes approaching the quantum-gravity scale. Since $M(x)$ is not an independent field but a quantity determined by the background geometry and the noncommutative structure, the violation arises purely from the effective modification of particle dynamics induced by spacetime noncommutativity.\\


Despite not considering any specific model of noncommutative gravity and rather discussing a model for \textit{probing} a noncommutative spacetime, it is important to notice some consistency with the general (model-independent) properties of noncommutative gravities: the first non-trivial correction always starts with $\Theta^2$ order \cite{Mukherjee:2006nd, Aschieri:2005yw, Alvarez-Gaume:2006qrw, Calmet:2006iz}.\\

Another point of view on the result \eqref{ncgeo1} is that it can be thought of as a toy model for a concrete realization of the so-called rainbow gravity \cite{Magueijo:2002xx}. Indeed, while in the original rainbow gravity theory different wave-lengths ``see'' different gravity, in our case, noncommutative corrections to geodesics are proportional to the mass of the probe particle. In particular, in our model, massless particles will \textit{not} see noncommutativity.\footnote{It is important to stress that the conclusion that massless particles are unaffected by the noncommutativity approach studied here applies to the scalar Klein-Gordon field considered here. Extending the analysis to gauge fields or fermions would, in fact, require studying their corresponding noncommutative wave equations and lies beyond the scope of the present work. } At the same time, the more massive the probe particle is, the stronger  noncommutativity it will feel. Note that due to \eqref{M2}, this is a general result that does not depend on the metric (a different metric could only either enhance or weaken this effect).\\

Finally, we conclude this discussion by pointing out some possible generalizations and future prospects for applications of the results presented in this paper. First, an obvious generalization is to use a more general NC algebra (here we basically used $[r, \theta]=i\lambda$), i.e., a more complicated twist than \eqref{rttwist}. The only restriction in our derivation is that the twist used must be Drinfeld (to guarantee the associativity of the $\star$-product) and that it is of Moyal-type in some coordinate system. This preferred coordinate system will define the so-called ``nice basis'' (see \cite{schenkel, Herceg:2025rll}  for more details). The second generalization stems from the choice of the metric \eqref{metric_spher}. Namely, we can keep the spherical symmetry, but allow for certain anisotropies \cite{Jacobson:2007tj}, meaning we can use metrics of the form
\begin{equation}\label{genmet}
ds^2=-Fdt^2+Gdr^2+H(d\theta^2 + \sin^2 \theta \, d\phi^2)
\end{equation}
where $F,G$ and $H$ are arbitrary functions of $r$ and $t$, and $FG\neq 1$.\footnote{Note that in our approach the corrections are obtained in a systematic way, which is rather different from the usual ``naive'' approach based on putting star-product between the metric coefficients. E.g., in our approach the results will differ from the corrections based on the following modification of \eqref{genmet}: $ds^2_\star=-Fdt^2+Gdr^2+H\star(d\theta^2 + \sin^2 \theta \, d\phi^2)$.} For the metric defined by \eqref{genmet}, most of the results of section IV (like \eqref{series}, \eqref{2nd}, and \eqref{4th}) would receive only a slight modification. More precisely, one only needs to ``re-scale'' the determinant of the metric as $g\mapsto \frac{FGH}{r^2}g$. This will enable us to apply our results to metrics that describe cosmological backgrounds \cite{Green:2014aga, Kantowski:1966te}, anisotropic situations \cite{Letelier:1980mxb, Cardoso:2019rvt, Herrera:1997plx, Mak:2001eb, C:2024cnk}, loop quantum gravity black holes \cite{Modesto:2009ve, Modesto:2005zm, Gingrich:2023fxu, Yang:2023gas, Cruz:2015bcj} and noncommutative black holes \cite{Gong:2023ghh, Juric:2025kjl, AraujoFilho:2026rdc, ref2}, to name a few.

%
%
%
%
%
%
\section*{Funding} This research was supported by the Croatian Science Foundation Project No. IP-2025-02-8625, \emph{Quantum aspects of gravity} and by the \emph{MZ2-Mobility of researchers grant} from the National Recovery and Resilience Plan of the Ministry of Science and Education of the Republic of Croatia. A.P. acknowledges the partial support of CNPq under the grant no.312842/2021-0 during the initial stages of this project.

\section*{Author contributions} All authors have equally contributed to this research. All authors have read
and agreed to the submitted version of the manuscript.

\section*{Data Availability Statement}
Data sharing is not applicable to this article, as no datasets were generated or analyzed during the current study.

\appendix

\section{Twisted Laplace-Beltrami operator}

The geometric definition of the Laplace-Beltrami operator acting on a scalar field $\Phi$ is given by
\ba\label{dAl}
\Box_g\Phi = * d\! *\! d \Phi .
\ea
While the exterior derivative $d$ carries the information on the differential structure, the metric structure is encoded in the Hodge star $*$, which is a map
$ *: \Omega^r \rightarrow \Omega^{n-r} $, and in the commutative case is explicitly given by
\ba\label{Hodge}
* \omega = \frac{\sqrt{g}}{r!(n-r)!} \omega_{\mu_1 \ldots, \mu_r} \epsilon^{\mu_1 \ldots \mu_r}_{~~~~~~~\nu_{r+1} \ldots \nu_{n}} dx^{\nu_{r+1}} \wedge\ldots\wedge dx^{\nu_{n}},
\ea
where $\omega = \frac{1}{r!} \omega_{\mu_1 \ldots, \mu_r} dx^{\mu_{1}} \wedge\ldots\wedge dx^{\mu_{r}}\in \Omega^r$. The metric information is not only in $g \equiv \det (g_{\mu\nu})$, but also in raising the indices of $\epsilon_{\mu_{1} \ldots \mu_{n}}$, which is normalized as $\epsilon_{{1} \ldots {n}} =1$.\\

From this discussion, we see that defining a noncommutative analogue of the Hodge $*$  introduces some variant of the noncommutative metric or, at least, gives a noncommutative generalization of the Laplace-Beltrami operator. Here we review how it is done for a noncommutative space defined via a Drinfeld twist, specifying to the simplest case of the Moyal space.\\

The differential properties of a smooth manifold $\mathcal{M}$ are captured in its \textbf{commutative} algebra of smooth functions $\mathcal{A}=(\mathcal{C}^{\infty}(\mathcal{M}), \cdot)$, where $\cdot$ stands for the commutative multiplication. Then, one takes as a definition of a noncommutative deformation of the manifold $\mathcal{M}$ a \textbf{noncommutative} algebra $\mathcal{A}_\star=(\mathcal{C}^{\infty}(\mathcal{M}), \star)$, where the $\star$-product is now some noncommutative product with respect to which the algebra $\mathcal{A}_\star$ is associative. One standard way to construct such a product is via the so-called Drinfeld twist
\begin{equation}\label{twist}
	\mathcal{F}=f^{A}\otimes f_{A},
\end{equation}
where $f$'s are some differential operators, and the inverse is $\mathcal{F}^{-1} = \bar{f}^A \otimes \bar{f}_A$. $\mathcal{F}$ and $\mathcal{F}^{-1}$ are acting on $\mathcal{C}^{\infty}(\mathcal{M})\otimes \mathcal{C}^{\infty}(\mathcal{M})$. In terms of the twist, the associative $\star$-product for any two elements of $\mathcal{C}^{\infty}(\mathcal{M})$ is given by $f\star g = m\circ \mathcal{F}^{-1} (f\otimes g)$, where $m : \mathcal{C}^{\infty}(\mathcal{M})\otimes \mathcal{C}^{\infty}(\mathcal{M}) \rightarrow \mathcal{C}^{\infty}(\mathcal{M})$ is the usual multiplication map. As we said above, instead of giving general definitions, we apply this to the simplest case of the Moyal noncommutative space. So, let $\Theta^{\mu\nu}$ be a constant antisymmetric matrix and $\left\{\partial_{\mu}\right\}$ be coordinate basis vector fields on $T\mathcal{M}$, then the Moyal twist and its inverse are defined as
\begin{equation}\label{Moyal_twist}
	\mathcal{F}=f^{A}\otimes f_{A} = e^{-\frac{i}{2}\Theta^{\mu\nu}\partial_{\mu}\otimes\, \partial_{\nu}},\ \mathcal{F}^{-1}=\bar{f}^{A}\otimes \bar{f}_{A} = e^{\frac{i}{2}\Theta^{\mu\nu}\partial_{\mu}\otimes\, \partial_{\nu}},
\end{equation}
where the exponent should be understood as formal series expansion. From this, one can easily read of $f$'s and $\bar{f}$'s, e.g.,
 \begin{equation}\label{barf}
\bar{f}^{A} = \frac{1}{n!}\left(\frac{i}{2}\right)^n \Theta^{\mu_1\nu_1}\cdots\Theta^{\mu_n\nu_n}\partial_{\mu_1}\cdots\partial_{\mu_n},\
\bar{f}_{A} = \partial_{\nu_1}\cdots\partial_{\nu_n}.
\end{equation}
Immediately, \eqref{Moyal_twist}  leads to the well-known expression for the Moyal $\star$-product
\begin{equation}\label{Moyal_star}
	(f\star g)(x) = m\circ \mathcal{F}^{-1} (f\otimes g) = e^{\frac{i}{2}\Theta^{\mu\nu}\partial^x_{\mu}\, \partial^y_{\nu}} f(x)g(y)|_{x=y} \equiv \sum\limits_{n=0}^{\infty}\frac{1}{n!}\left(\frac{i}{2}\right)^n\Theta^{\mu_1\nu_1}\cdots\Theta^{\mu_n\nu_n} \left(\partial_{\mu_1}\cdots\partial_{\mu_n}f(x)\right) \Big(\partial_{\nu_1}\cdots\partial_{\nu_n}g(x)\Big) .
\end{equation}
Of course, this means that the Moyal twist defines a noncommutative space generated by the ``noncommutative coordinates''
\begin{equation}
[x^{\mu} \stackrel{\star}{,} x^{\nu}]=x^{\mu}\star x^{\nu}-x^{\nu}\star x^{\mu}=i \Theta^{\mu\nu}.
\end{equation}
The twist \eqref{twist} enables us to define NC geometric objects like Lie derivative  $\pounds^{\star}$, connection $\hat{\nabla}$, curvature tensor $\hat{R}$ and torsion $\hat{T}$, as well as allows us to deform the notion of the symmetries for the noncommutative setting that are now encoded in the $\star$-Lie algebra of vector fields and the corresponding Hopf algebra (for more details see \cite{Herceg:2023zlk,
Herceg:2023pmc, Aschieri:2005zs, Aschieri:2005yw, Aschieri:2009qh, schenkel, Aschieri:2017ost, Aschieri:2020yft} ). Specifying to the Moyal twist \eqref{Moyal_twist}, we can define a general vector field $v$ as
\begin{equation} v= v^\mu_\star \star \partial_\mu = v^\mu\partial_\mu, \end{equation}
and, in particular, $v^\mu=v^\mu_\star$ (which is true in the Moyal case, but false in the case of a general twist). In the same way, any one-form can be written as
\begin{equation}\omega= \omega^\star_\mu \star d x ^\mu = \omega_\mu d x^\mu, \end{equation}
with $\omega_\mu^\star = \omega_\mu$. Defining the twisted wedge product, $\wedge_\star$, analogously to the $\star$-product, i.e., for any two 1-forms,
\begin{equation}\label{wedge}
\omega_1 \wedge_\star \omega_2 := \bar{f}^A (\omega_1) \wedge \bar{f}_A (\omega_2),
\end{equation}
we can construct by induction the space of twisted $k$-forms, $\Omega^k_\star$ with the usual exterior derivative, $d : \Omega^k_\star \rightarrow \Omega^{k+1}_\star$, satisfying the undeformed graded Leibnitz rule.\\

As an important demonstration, let us find the explicit form for a homogeneous $k$-form $\omega = f\star dx^{\mu_1}\wedge_\star \cdots \wedge_\star dx^{\mu_k}\in \Omega^k_\star$. For a basis 2-form $dx^{\mu_1}\wedge_\star dx^{\mu_2}$, we have from \eqref{wedge}
\begin{equation}\label{wedge2}
dx^{\mu_1}\wedge_\star dx^{\mu_2} := \bar{f}^A (dx^{\mu_1}) \wedge \bar{f}_A (dx^{\mu_2}) \equiv dx^{\mu_1}\wedge dx^{\mu_2},
\end{equation}
where we used that\footnote{This trivially follows from the fact that $\partial_\mu$ acts on $dx^\nu$ as a commutative Lie derivative $\pounds_{\partial_\mu}$. But for any form $\omega$ and any vector field $X$, one has $\pounds_{X} \omega = (d\circ i_X + i_X \circ d)\omega$, which gives $\pounds_{\partial_\mu} (dx^\nu) = 0$, so only $A=0$ terms in \eqref{barf}, $\bar{f}^0 = \bar{f}_0 = 1$, contribute.} $\bar{f}^A (dx^{\mu}) = \bar{f}_A (dx^{\mu})=dx^{\mu}$. Then, by repeated application of \eqref{wedge}, we get
\begin{equation}\label{wedge3}
\omega = f\star dx^{\mu_1}\wedge_\star \cdots \wedge_\star dx^{\mu_k} = f dx^{\mu_1}\wedge \cdots \wedge dx^{\mu_k} \equiv dx^{\mu_1}\wedge_\star \cdots \wedge_\star dx^{\mu_k} \star f .
\end{equation}

After introducing the main ingredients of Moyal differential geometry, we would like to have a noncommutative generalization of the Hodge star as a right $\mathcal{A}_\star$-linear map, $*^\mathcal{F} : \Omega^r_\star \rightarrow \Omega^{n-r}_\star$, such that for any $\omega\in \Omega^*_\star$ and any $f\in \mathcal{A}_\star$
\begin{equation}\label{NCHodge}
*^{\mathcal{F}} (\omega \star f) = *^\mathcal{F}(\omega) \star f .
\end{equation}
Essentially, we want to consistently deform the commutative right module map, $*$, into the noncommutative right module map, $*^{\mathcal{F}}$. The recipe is well known \cite{Aschieri:2017ost, 
Aschieri:2020yft} as the \textit{quantization map} and it is given by
\begin{equation}\label{NCHodge1}
\omega \mapsto *^\mathcal{F}(\omega) = \overline{f}^A_1 (* (S(\overline{f}^A_2) \overline{f}_A(\omega))),
\end{equation}
where \( S \) is the antipode, and \[\Delta \overline{f}^A =: \overline{f}^A_1 \otimes \overline{f}^A_2\] is the co-product. For the Moyal case, antipode and co-product are given by the standard rules for a Lie algebra:
\begin{equation}\label{SDelta}
S(1)=1,\ \ S(\partial_\mu) = -\partial_\mu ,\ \ \Delta(1)=1\otimes 1,\ \ \Delta(\partial_\mu) = \partial_\mu \otimes 1 + 1 \otimes \partial_\mu .
\end{equation}
By the homomorphism property of $\Delta$, we have
\begin{equation}\label{Delta}
\Delta(\partial_{\mu_1}\cdots \partial_{\mu_k}) = (\partial_{\mu_1} \otimes 1 + 1 \otimes \partial_{\mu_1} ) \cdots (\partial_{\mu_k} \otimes 1 + 1 \otimes \partial_{\mu_k} ).
\end{equation}
From \eqref{wedge3}, we know that only terms with $A=0$ will contribute to \eqref{NCHodge1}, leading to, with the $k$-form $\omega$ as above,
\begin{equation}\label{NCHodge2}
*^\mathcal{F}(\omega) = \overline{f}^0_1 (* (S(\overline{f}^0_2) \overline{f}_0(dx^{\mu_1}\wedge \cdots \wedge dx^{\mu_k})))\star f \equiv \left(* (dx^{\mu_1}\wedge \cdots \wedge dx^{\mu_k})\right)\star f,
\end{equation}
where we used that, from \eqref{barf}, $\bar{f}^0 = \bar{f}_0 = 1$. We stress that, on the right hand side of \eqref{NCHodge2}, the $*$ is the usual commutative Hodge star \eqref{Hodge}.\\

Now we are ready to define and calculate the noncommutative analogue, $\hat{\square}_g$, of the Laplace-Beltrami operator \eqref{dAl}. The natural definition is
\begin{equation}\label{NCdAl}
\hat{\square}_g\Phi \equiv *^\mathcal{F} d*^{\mathcal{F}}d\Phi ,
\end{equation}
where $\Phi$ is a noncommutative (scalar) field. Using the result \eqref{NCHodge2}, we have
\ba\label{dPhi}
d\Phi &=&(\partial_\mu\Phi)\star d x ^\mu = \partial_\mu \Phi\, d x^\mu \equiv d x ^\mu\star(\partial_\mu\Phi), \nonumber\\
*^\mathcal{F}(d\Phi) &=& *^\mathcal{F}(d x ^\mu \star \partial_\mu \Phi) = \frac{1}{3!} dx^{\nu_1}\wedge dx^{\nu_2} \wedge dx^{\nu_3} \left(\sqrt{g}~g^{\mu\nu} \epsilon_{\nu\nu_1\nu_2\nu_3}\right)\star \partial_\mu\Phi .
\ea
Taking exterior derivative
\begin{equation}
d(*^\mathcal{F}d\Phi)=\frac{1}{3!} dx^\tau \wedge_\star dx^{\nu_1}\wedge_\star dx^{\nu_2} \wedge_\star dx^{\nu_3} \star \left[\partial_\tau \left[\left(\sqrt{g} g^{\mu\nu}\epsilon_{\nu\nu_1\nu_2\nu_3}\right)\star \partial_\mu\Phi\right]\right],
\end{equation}
we get
\begin{eqnarray}
*^\mathcal{F} d*^{\mathcal{F}}d\Phi &=& *^\mathcal{F} \left( dx^\tau \wedge_\star dx^{\nu_1}\wedge_\star dx^{\nu_2} \wedge_\star dx^{\nu_3}\right) \star \left[\partial_\tau \left[\left(\frac{1}{3!}\sqrt{g} g^{\mu\nu}\epsilon_{\nu\nu_1\nu_2\nu_3}\right)\star \partial_\mu\Phi\right]\right] \nonumber \\
&=& \frac{1}{\sqrt{g}}\star \left[\partial_\nu \left[\left(\sqrt{g}~g^{\mu\nu}\right)\star \partial_\mu\Phi\right]\right],
\end{eqnarray}
and therefore, finally
\begin{equation}\label{NCdAlfinal}
\hat{\square}_g\Phi = \frac{1}{\sqrt{g}}\star \left[\partial_\nu \left[\left(\sqrt{g}~g^{\mu\nu}\right)\star \partial_\mu\Phi\right]\right].
\end{equation}

Notice that \eqref{NCdAlfinal} is a rather unambiguous and not entirely trivial result in this approach. For instance, naively one might expect that one should have $\star$-product between all the functions entering the expression. Instead, we see that the combination $\sqrt{g}~g^{\mu\nu}$ appears with the usual, commutative, multiplication.

\section{On the algebraic $\star$-inverse}
As mentioned in section III, the central object to calculated is the algebraic $\star$-inverse of $1/\sqrt{g}$, i.e., the function $\sqrt{g}_\star$ defined by 
\begin{equation}
\sqrt{g}_\star \star \frac{1}{\sqrt{g}}=1.
\end{equation}
The purpose of this appendix is to present some properties of the algebraic $\star$-inverse in general (first subsection), calculate the general formula for a specific type of noncomutativity (second subsection), and evaluate this formula for explicit examples up to fifth order in the NC parameter (third subsection).
Here we are using the metric \eqref{metric_spher} and the NC deformation defined by the twist \eqref{rttwist}, meaning that the noncommutivity is of the $(r-\theta)$-type, and that the determinant of the metric $g$ (and its algebraic inverse) can always be written as a product of single variable dependent functions, i.e. $\sqrt{g}=f(r)g(\theta)$. Therefore, in the second subsection of this appendix, we will adopt an obvious change of notation: $r\rightarrow x$, $\theta\rightarrow y$ and $\mathcal{F}\rightarrow \exp{(\frac{i\lambda}{2}\partial_{x}\wedge\partial_y)}$.

\subsection*{1. Proposition on Even Functions}

\textbf{Proposition:}
Let \( f(x;\Theta) \in C^\infty(\mathbb{R}^n) \) be even in \( \Theta \), i.e., \( f(x;-\Theta) = f(x;\Theta) \).
Also let \( \star_\Theta \) be some Moyal-type associative $\star$-product
(meaning \( (f \star_\Theta g) \star_\Theta h = f \star_\Theta (g \star_\Theta h) \equiv f \star_\Theta g \star_\Theta h \) as in the previous appendix. Then \( f_\star^{-1} \), defined by \( f_\star^{-1} \star_\Theta f = 1 \),
does not depend on odd degrees of \( \Theta \).\\

\textbf{Proof:}
By associativity and uniqueness of \( 1 \), we have
\[
f_\star^{-1} \star_\Theta f = 1 = f \star_\Theta f_\star^{-1},
\]
i.e., \( f_\star^{-1} \) is a left/right inverse.
Then by the condition that $f$ is even in $\Theta$, we have
\[
f_\star^{-1}(x;-\Theta) \star_{-\Theta} f(x;-\Theta) = f(x;\Theta) \star_\Theta f_\star^{-1}(x;-\Theta),
\]
where we used that for any two functions, $f,g$, one has $f \star_{-\Theta} g = g \star_\Theta f$ (this easily follows from the anti-symmetry of $\Theta^{\mu\nu}$).
By uniqueness of the inverse, we must have
\[
f_\star^{-1}(x;-\Theta) = f_\star^{-1}(x;\Theta),
\]
i.e., \( f_\star^{-1}(x,y;\Theta) \) also depends only on even powers of \( \Theta \).
Also, for any real function $f(x;\Theta)$, we have:
\[
\overline{f_\star^{-1}(x;\Theta) \star_{\Theta} f(x;\Theta)} = f(x;\Theta) \star_\Theta \overline{f_\star^{-1}(x;\Theta)},
\]
where we used that for any two functions, $f,g$, one has $\overline{f \star_{\Theta} g} = \bar{g} \star_\Theta \bar{f}$. Again, by the uniqueness of the inverse, we have that $f_\star^{-1}(x;\Theta)$ is also real. In particular, if $f(x;\Theta)$ is real and even in $\Theta$,
\( f_\star^{-1} \)  is also real and depends only on even degrees of \( \Theta \).

\subsection*{2. $\star$-Inverse of $f(x)g(y)$}

Let us study the \( \star \)-inverse of \( f(x)g(y) \), where \( f \) and \( g \) are \( C^\infty \), \( \Theta \)-independent functions. What follows is a purely perturbative construction. We assume the existence of the algebraic $\star$-inverse as a formal power series in the noncommutativity parameter and make no claim regarding the convergence or radius of convergence of the full series. 
Rename \( f(x)g(y) \mapsto e^{f(x)} e^{g(y)} \).
We look for \( (f(x)g(y))_\star^{-1} \) in the form:
\[
(f(x)g(y))_\star^{-1} = e^{-f} e^{-g} \sum_{n=0}^\infty \lambda^{2n} F_n(x,y).
\]
\textbf{Notation:}
\( f = f(x) \) and \(g = g(y)\), so we will use the following notation \( f' = \dd_{x}{f} \), \( g' = \dd_{y}{g} \), etc.\\

The \( \star \)-product is defined as:
\[
f \star g = e^{{i\lambda}(\overleftarrow{\partial}_x \overrightarrow{\partial}_y - \overleftarrow{\partial}_y \overrightarrow{\partial}_x)} f(x,y) g(x,y),
\]
we use $\lambda$ as a deformation parameter instead of $\Theta$ to avoid the factor of $\frac{1}{2}$ in the $\star$-product, as in \eqref{Moyal_star}, and we must have:
\[
\sum_{n=0}^\infty \lambda^{2n} \bigl( F_n(x,y) e^{-f} e^{-g} \bigr) \star \bigl( e^{f} e^{g} \bigr) = 1 .
\]
This expands as:
\ba\label{expansion}
\sum_{n,s,\ell} \lambda^{2n+s+\ell} \frac{i^s (-i)^\ell}{s!\, \ell!} \partial_x^{(s)} \partial_y^{(\ell)} \bigl( F_n(x,y) e^{-f} e^{-g} \bigr) \cdot \partial_x^{(s)} e^{f} \cdot \partial_y^{(\ell)} e^{g} = 1 .
\ea
\\

\textbf{Evaluating terms in orders of $\lambda$:}

Let $(n,s,\ell)$ be a set of degrees in \eqref{expansion}. Then we have up to $\mathcal{O}(\lambda^6)$
\begin{enumerate}
\item[$\lambda^0$,] \( (0,0,0) \): \( F_0 = 1 \).
\item[$\lambda^1$,] \( (0,1,0) + (0,0,1) =0 \), which is consistent with the general result from the proposition.
\item[$\lambda^2$,] \( (1,0,0) + (0,1,1) + (0,2,0) + (0,0,2) \):
\[
F_1 + f'' g'' = 0 \Rightarrow F_1 = (i)^2 f'' g'' .
\]
\item[$\lambda^3$,] \( (1,1,0)+ (1,0,1)+ (0,3,0)+ (0,2,1)+ (0,1,2)+ (0,0,3) =0 \), again, consistency check gives 0.
\item[$\lambda^4$,] \( (2,0,0)+ (1,2,0)+ (1,1,1)+ (1,0,2)+ (0,4,0)+ (0,3,1)+(0,2,2) + (0,1,3)+ (0,0,4):\) 
\[F_2 - 2\frac{1}{4!}f^{(4)}g^{(4)} + \frac{1}{2!}\bigl( (f'')^2 g^{(4)} + f^{(4)} (g'')^2 \bigr) =0 \quad \Longrightarrow \quad 
F_2 = 2\frac{1}{4!} f^{(4)} g^{(4)} - \frac{1}{2!} \bigl( (f'')^2 g^{(4)} + f^{(4)} (g'')^2 \bigr).
\]
\end{enumerate}

\textbf{So we get:}
\ba\label{expansion1}
\bigl( e^{f} e^{g} \bigr)_\star^{-1} =
e^{-f} e^{-g} \left( 1 + 2\Bigl( \frac{(i\lambda)^2}{2!} f'' g''  +\frac{(i\lambda)^4}{4!} f^{(4)} g^{(4)} \Bigr) - \frac{(i\lambda)^4}{2!} \bigl( (f'')^2 g^{(4)} + f^{(4)} (g'')^2 \bigr) + \mathcal{O}(\lambda^6)
\right).
\ea

\subsection*{3. Explicit Case}

A natural illustrative case to consider is \( e^{f} e^{g} = x^{-n} \sin^{-k} y \), i.e., \( f = -n\ln x \), \( g = -k \ln \sin y \), as for appropriate choices of coordinates and the parameter $k$, this reduces to familiar quantities. For the derivatives we have
\[
f^{(s)} = (-1)^s n \frac{(s-1)!}{x^s}, \quad s > 0 ,
\]
\[
g'' = \frac{k}{\sin^2 y}, \quad g^{(4)} = \frac{2k}{\sin^4 y}(2\cos^2 y + 1).
\]

Then, using this in \eqref{expansion1}, we have up to \( \mathcal{O}(\lambda^6) \):
\begin{eqnarray}
(x^{-n} \sin^{-k} y)_\star^{-1} &=&
x^n \sin^k y \left( 1 - \lambda^2 nk \frac{1}{x^2 \sin^2 y} + \frac{\lambda^4}{4!} 2 \cdot 3! \cdot 2nk \frac{1}{x^4 \sin^4 y} (2\cos^2 y + 1) \right. \nonumber \\
&&\left. + \frac{\lambda^4}{2} \left( n \frac{3! k^2}{x^4 \sin^4 y} + \frac{n^2 \cdot 2k}{x^4 \sin^4 y} (2\cos^2 y + 1) \right) + \mathcal{O}(\lambda^6) \right) \nonumber \\
&=& x^n \sin^k y \left( 1 - \lambda^2 nk \frac{1}{x^2 \sin^2 y} \right. \nonumber\\
 &&\left. + \lambda^4 nk \frac{3k+(1-n)(2\cos^2 y + 1)}{x^4 \sin^4 y} + \mathcal{O}(\lambda^6) \right). \nonumber
\end{eqnarray}

For the special case of $n=2$ and $k=1$ (which corresponds to $\sqrt{g}=r^2 \sin \theta$, compare this to \eqref{g_expension}), we have
\begin{eqnarray}\label{4th_xy}
\left(\frac{1}{x^{2} \sin y}\right)_\star^{-1} = x^2 \sin y \left( 1 -  \frac{2\lambda^2}{x^2 \sin^2 y}  + \frac{4\lambda^4}{x^4 \sin^2 y} + \mathcal{O}(\lambda^6) \right). \nonumber
\end{eqnarray}

\section{Nonrelativistic limit or the correspondence between Schrodinger equation and Newton's second law}\label{Nonrelativistic}

Here we give a complete standard argument for our approach in the non-relativistic case. This should clarify some steps in our relativistic generalization developed in Section \ref{sec:commutative}.\\

Let $\psi$ satisfy the Schrodinger equation
\ba\label{Schrodinger}
i \hbar\partial_t \psi = H(\hat{\vec{x}},\hat{\vec{p}})\,\psi \equiv \left( - \frac{\hbar^2}{2m} \nabla^2 + V(\vec{x}) \right) \psi(\vec{x})
\ea
and write \( \psi = A e^{\frac{i}{\hbar} S} \). At the moment, we are not making any assumption neither about $A$, nor about $S$. Let us evaluate $\langle \psi|\hat{\vec{p}}|\psi\rangle$:
\ba\label{everage}
\langle \psi|\hat{\vec{p}}|\psi\rangle = \frac{\hbar}{i}\int \left( \bar{A} e^{-\frac{i}{\hbar} S} \vec{\nabla}\left(A e^{\frac{i}{\hbar} S}\right) \right) dV = \int \left( |A|^2 \vec{\nabla}S + \frac{\hbar}{i}\bar{A}\vec{\nabla}A \right) dV .
\ea
Now we assume that $\psi$ is a quasi-classical state, i.e., in \eqref{everage} one can neglect the second term compared to the first one\footnote{This essentially means that we are assuming that $S = S_0 + \hbar S_1 + \ldots$, $A = A_0 + \hbar A_1 + \ldots$ and that the higher order terms in $\hbar$ (and their gradients) are negligible in the quasi-classical limit. This is done in Section \ref{sec:commutative} for the general case.},
\ba\label{qc_everage}
\langle \psi|\hat{\vec{p}}|\psi\rangle \overset{``\hbar\rightarrow 0"}{\longrightarrow} \int  |A|^2 \vec{\nabla}S  dV .
\ea
This shows that, in this limit, we have
\ba
\hat{\vec{p}}\psi \longrightarrow (\vec{\nabla}S ) \psi,
\ea
as long as $\psi$ is quasi-classical\footnote{Formally, this follows directly from the following consideration:
\[
\frac{\hbar}{i}\vec{\nabla}\psi = (\vec{\nabla}S ) \psi - i\hbar (\vec{\nabla}\ln A ) \psi = (\vec{\nabla}S ) \psi + \mathcal{O}(\hbar),
\]
i.e., $\hat{\vec{p}} = \vec{\nabla}S +\mathcal{O}(\hbar)$ on $\psi = A e^{\frac{i}{\hbar} S}$.}. In order to better understand $S$, we plug $\psi = A e^{\frac{i}{\hbar} S}$ in \eqref{Schrodinger}. We have, in the same limit,
\ba\label{HJ_class}
0 = -i \hbar\partial_t \psi + H(\hat{\vec{x}},\hat{\vec{p}})\,\psi = \partial_t S + H({\vec{x}},{\vec{\nabla}S}) \psi(\vec{x})+\mathcal{O}(\hbar).
\ea
As $\psi \ne 0$, we conclude that, in the leading order
\ba\label{HJ_class1}
\partial_t S + H({\vec{x}},{\vec{\nabla}S})  = 0,
\ea
i.e., $S$ satisfies the classical Hamilton-Jacobi equation and hence it must be that $\vec{\nabla}S = \vec{p}$, which is consistent with the previous discussion. Assuming that $H({\vec{x}},{\vec{p}})= \frac{\vec{p}^2}{2m} + V(\vec{x})$, as in \eqref{Schrodinger} (and using $\vec{p} = m\dot{\vec{x}}$), we easily derive the equation for non-relativistic ``geodesics'', which is, of course, nothing but the Newton's second law
\ba\label{Newton}
\frac{d \vec{p}}{dt} = \frac{d}{dt} \vec{\nabla} S = \partial_t\vec{\nabla}S + (\dot{\vec{x}}\cdot \vec{\nabla})\vec{\nabla}S = \vec{\nabla}\left( \frac{\partial S}{\partial t} + \frac{(\vec{\nabla}S)^2}{2m} \right) \overset{\eqref{HJ_class1}}{\equiv} - \vec{\nabla}V(\vec{x}).
\ea


\end{document}